\documentclass{PoS}
\usepackage{multirow}

\title{On the nature of an excited state}

\ShortTitle{}

\author{\speaker{Beno\^\i t Blossier}\\
        Laboratoire de Physique Th\'eorique, CNRS, Univ. Paris-Sud et Universit\'e Paris-Saclay, B\^atiment 210, F-91405 Orsay Cedex, France\\
        E-mail: \email{benoit.blossier@th.u-psud.fr}}

\author{Antoine G\'erardin\\
        PRISMA Cluster of Excellence and Institut f\"ur Kernphysik, University of Mainz, Becher-Weg 45, 55099 Mainz, Germany\\
        }

\abstract{In many lattice simulations with dynamical quarks, radial or orbital 
excitations of hadrons lie near multihadron thresholds: it makes the 
extraction of excited states properties more challenging and can introduce 
some systematics difficult to estimate without an explicit computation of 
correlators using interpolating fields strongly coupled to multihadronic 
states. In a recent study of the strong decay of the first radial excitation 
of the $B^*$ meson, this issue has been investigated and we have clues that 
a diquark interpolating field $\bar{b} \gamma^i q$ is very weakly coupled to a 
$B \pi$ $P$-wave state while the situation is quite different if we consider an 
interpolating field of the kind $\bar{b} \nabla^i q$, where $\vec{\nabla}$ is a 
covariant derivative: those statements are based on examining the charge 
density distribution.
}

\FullConference{34th annual International Symposium on Lattice Field Theory\\
		24-30 July 2016\\
		University of Southampton, UK}

\begin{document}

\section{Lattice estimate of the coupling $g_{B^{*\prime}B\pi}$}

\begin{figure}[t]
\begin{center}
\begin{tabular}{cc}
\includegraphics*[height=5cm,width=7cm]{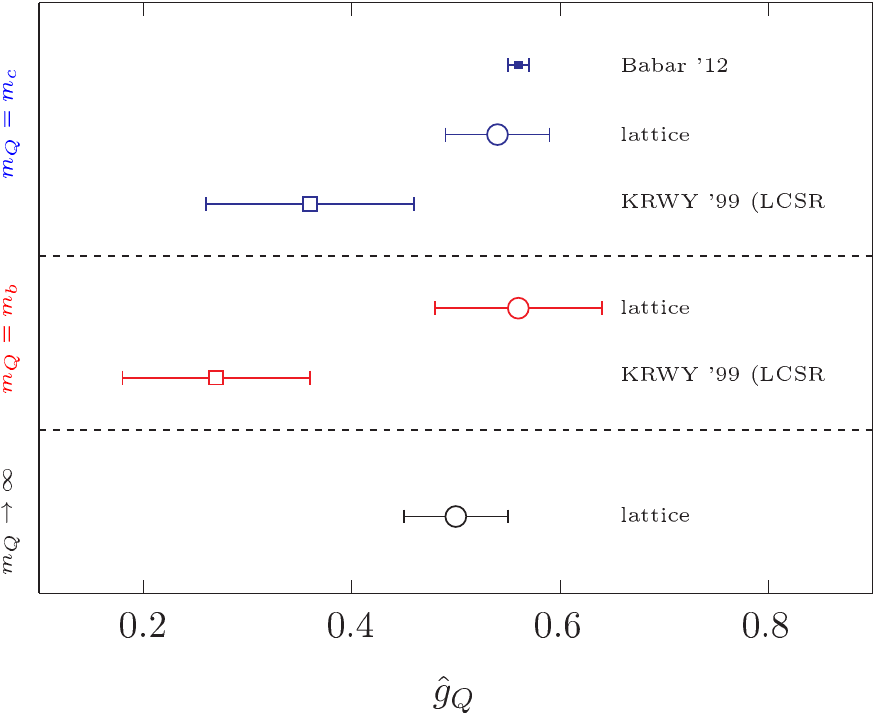}
&\includegraphics*[width=6cm,height=4cm]{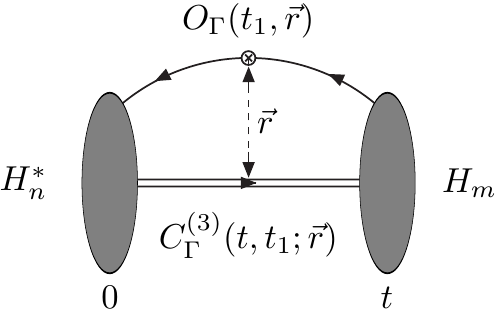}
\end{tabular}
\end{center}
\caption{\label{figdensity} Experimental measurement, lattice 
computations and sum rules estimates of $\hat{g}_c$, $\hat{g}_b$ and $\hat{g} \equiv \hat{g}_\infty$
(left); three-point correlation function computed to extract the density distribution $f^{(mn)}_{\Gamma}(\vec{r})$ (right).}	
\end{figure}
Questions have been raised on the poor handling of excited states in 
analytical computations of quantities directly related to the dynamics at work in
strong interaction. For instance it 
has been argued that the light-cone sum rule determination of the 
$g_{D^{*}D\pi}$ coupling, which parametrises the $D^{*} \to D \pi$ decay, 
likely fails to reproduce the 
experimental measurement unless one explicitly includes
a (negative) contribution from the first radial excited $D^{(*)'}$ state on the 
hadronic side of the three-point Borel sum rule~\cite{BecirevicVP}. Comparison 
with sum rules is of particular importance because the heavy mass dependence 
of $\hat{g}_Q\equiv \frac{g_{H^*H\pi}f_\pi}{2\sqrt{m_H m_{H^*}}}$ deduced from 
recent lattice simulations~\cite{OhkiPY} and 
experiment~\cite{GodangIM} seems much weaker than expected from analytical 
methods~\cite{KhodjamirianHB}, as shown in the left panel of Fig.~\ref{figdensity}. We have proposed to test the hypothesis of $g_{D^{*'}D\pi}<0$ by doing a direct computation of the coupling $g_{B^{*'}B\pi}$, in the static limit of Heavy Quark Effective Theory (HQET), assuming the smoothness of results in $1/m_b$, $1/m_c$ in order to qualitatively relate charm and bottom regions \cite{BlossierQMA}. The transition amplitude of interest $T^{mn\,\mu}=\langle B_m(p) |{\cal A}^{\mu}| B_n^{*}(p^{\prime},\lambda) \rangle$ is parametrized by 
\begin{eqnarray}
\nonumber
T^{(mn)\,\mu}&=& 2m_{B_n^{*}}A^{(mn)}_0(q^2) \frac{\epsilon(p^{\prime},\lambda)\cdot q}{q^2} q^\mu +(m_{B_m} + m_{B_n^*}) A^{(mn)}_1(q^2)\left(\epsilon^{\mu}(p^{\prime},\lambda) -  \frac{\epsilon(p^{\prime},\lambda) \cdot q}{q^2} \, q^\mu \right)\\
&+&A^{(mn)}_2(q^2) \frac{\epsilon(p^{\prime},\lambda) \cdot q}{m_{B_m} + m_{B_n^*}} \left[(p+p^{\prime})^\mu + \frac{m^2_{B_m} - m^2_{B_n^*}}{q^2}q^\mu \right] \,,
\end{eqnarray}
with ${\cal A}^{\mu}(x) = \overline{d}(x) \gamma^\mu \gamma^5 u(x)$ and $q=p-p^\prime$. Taking the divergence of the current $q_\mu {\cal A}^\mu$, using the Partially Conserved Axial Current (PCAC) relation, the LSZ reduction formula and $\sum_{\lambda} \epsilon_{\mu}(k,\lambda) \, \epsilon^{*}_{\nu}(k,\lambda) =  - g_{\mu\nu} + \frac{k_{\mu}k_{\nu}}{m^2}$, we are left with
\begin{equation}
g_{B_n^{*} B_m \pi}    =  \frac{ 2 \, m_{B_n^{*}}A^{(mn)}_0(0)}{f_{\pi}}  \,,\quad
A^{(mn)}_0(q^2) =  - \sum_{\lambda} \frac{  \langle B_m(p) |q_{\mu} {\cal A}^{\mu} | B_n^{*}(p',\lambda) \rangle  }{ 2 m_{B_n^*} \, q_i}  \, \epsilon^{*}_i(p^{\prime}, \lambda).
\end{equation}
Back to the $x$ space, we have
\begin{equation}
A^{(mn)}_0(q^2=0) = - \frac{q_0}{q_i} \int d^3 r\, f_{\gamma_{0} \gamma_5}^{(mn)}(\vec{r}) \, e^{i \vec{q}\cdot {\vec r}} + \int d^3 r \, f_{\gamma_{i} \gamma_5}^{(mn)}(\vec{r}) \, e^{i \vec{q}\cdot {\vec r}},
\end{equation}
where we have introduced density distributions $f^{(mn)}_{\gamma_\mu\gamma_5}(\vec{r})$ that are defined in terms of 2-pt and 3-pt HQET correlation functions: the latter is sketched in the right panel of Fig.~\ref{figdensity}. Analysing a set of ${\rm N_f}=2$ CLS ensembles made with ${\cal O}(a)$ improved Wilson-Clover fermions, whose parameters are collected in Table~\ref{tabsim}, we extract the coupling $g_{B^{*\prime}B\pi}$ from the contributions in Fourier space
\begin{table}[t]
\begin{center}
\begin{tabular}{|c|c|c|c|c|c|c|c|}
\hline
	lattice	&	$\quad\beta\quad$	&	$(L/a)^3\times T/a$ 		& $\kappa$ & $a[\rm fm]$	&	$m_{\pi}[\rm MeV]$	& $Lm_{\pi}$ &		$\{R_1,R_2,R_3\}$\\
\hline
	A5	&	$5.2$ 			&  	$32^3\times64$	&	$0.13594$&	$0.075$  	& 	$330$	&4	&  $\{15,60,155\}$ \\ 
	B6	&					&	$48^3\times96$	&	$0.13597$&		   	&	$280$	&5.2	&\\ 
\hline
	
	D5	&	$5.3$			&	$24^3\times48$	& $0.13625$&	 	$0.065$  	& 	$450$	&3.6	&  $\{22,90,225\}$	 \\  
	E5	&					&	$32^3\times64$	& $0.13635$&	 		  	& 	$440$	&4.7	&	 \\  
	F6	&		 	 		& 	$48^3\times96$	& $0.13638$&	 			& 	$310$	&5	&	 \\    
\hline
	
	N6	&	$5.5$ 			&	$48^3\times96$	& $0.13667$&	 	$0.048$  	& 	$340$	&4	&  $\{33,135,338\}$	 \\  
	
\hline
\hline
	Q1	&	$6.2885$ 			&	$24^3\times48$	&$0.13849$&		$0.06$  	& 	$\times$	& $\times$&\{22,90,225\}\\
Q2	&	$6.2885$ 			&	$32^3\times64$	&	$0.13849$&	$0.06$  	& 	$\times$	&  $\times$&\\ 
\hline
 \end{tabular} 
\end{center}
\caption{Parameters of the simulations: bare coupling $\beta = 6/g_0^2$, lattice resolution, hopping parameter $\kappa$, lattice spacing $a$ in physical units and pion mass. The smeared quark field are defined as $\psi^{(i)}_l (x)= (1+ \kappa_G a^2 \Delta)^{R_i} \psi_l (x)$ where $\kappa_G=0.1$ and $\Delta$ is the covariant Laplacian made with APE-blocked links. Sets D5, Q1 and Q2 are not used to extrapolate our results at the physical point: they are used to study finite volume and quenching effects. The quark mass for Q1 and Q2 is tuned to the strange quark mass.}
\label{tabsim}
\end{table}
\begin{eqnarray}
\nonumber
\mathcal{M}_i (q^2_{\rm max} - \vec{q}\,^2)& =& 4\pi \int_0^{\infty} \mathrm{d}r \ r^2 \, \frac{\sin(|\vec{q}|r)}{|\vec{q}| r}  f_{\gamma_{i} \gamma_5}^{(12)}(\vec{r}),\\
\nonumber
\frac{q_0}{q_i} \mathcal{M}_0 (q^2_{\rm max} - \vec{q}\,^2) &=& - q_0 4 i \pi  \int_{0}^{\infty} \mathrm{d} r_{\parallel}  \int_{0}^{\infty} \mathrm{d} r_{\perp} \, r_{\perp} \, f^{(12)}_{\gamma_0\gamma_5}(r_{\parallel},r_{\perp}) \, \frac{\sin(|\vec{q}| \, r_{\parallel})}{|\vec{q}|},
\end{eqnarray}
\begin{equation}
A^{(12)}_0(q^2)=-\frac{q_0}{q_i} \mathcal{M}_0 (q^2_{\rm max} - \vec{q}\,^2) + \mathcal{M}_i (q^2_{\rm max} - \vec{q}\,^2).
\end{equation}
Extrapolation of $A^{(12)}_0(q^2=0)$ at the physical point has been performed using the formula
\begin{equation}
A^{(12)}_0(0,m^2_\pi)=D_0 + D_1 a^2 + D_2 m^2_\pi/(8\pi f^2_\pi).
\end{equation}
Our result reads finally
\begin{equation}
A^{(12)}_0(0) =  -0.173(31)_{\rm stat}(16)_{\rm syst}, \quad g_{B^{*\prime} B\pi} =  -15.9(2.8)_{\rm stat}(1.4)_{\rm syst},
\end{equation}
while a computation done in the quenched approximation and at the strange mass gives $A^{(12)}_0(0) = -0.143(14)$. As seen in Table~\ref{tab:ff_comp} a comparison with two quark models, that are appealing in the heavy quark limit, draws the conclusion of a qualitative agreement in the fact that $q_0/q_i {\cal M}_0$ dominates over ${\cal M}_i$ in their respective contribution to $A^{(12)}_0(q^2=0)$ and it explains the negative sign of $g_{B^{*\prime}B\pi}$.
\begin{table}[t] 
\begin{center}
\begin{tabular}{|c|cc|cc|cc|}
	\hline
	&	\multicolumn{2}{c|}{Lattice}	&	\multicolumn{2}{c|}{BT}	&	\multicolumn{2}{c|}{Dirac} \\
	\hline 
	$q^2$& 	$q^2_{\rm max}$ & $0$ & 	$q^2_{\rm max}$ & $0$ & 	$q^2_{\rm max}$ & 	$0$ \\
	\hline 
	$q_{0} \mathcal{M}_{0}(q^2) / q_i$ & $0.402(54)_{\rm stat}(27)_{\chi}$ & $0.237(27)_{\rm stat}(28)_{\chi}$ &0.252&0.173&0.219&0.164\\  
	$\mathcal{M}_{i}(q^2)$ & $-0.172(16)_{\rm stat}(6)_{\chi}$ & $0.064(9)_{\rm stat}(13)_{\chi}$&-0.103&0.05&-0.223&-0.056  \\
	\hline
 \end{tabular} 
\end{center}
\caption{Lattice and quark models results for the spatial and time contributions to $A^{(12)}_0(q^2)$ at the kinematical points $q^2_{\rm max}$ and 0 \cite{LeYaouancprivate}.
Left panel: Extrapolated lattice results using the fit formula (1.5): the first error is statistical and the second error include the systematics from the chiral extrapolation.
Middle panel: Bakamjian-Thomas (BT) with Godfrey-Isgur potential, obtaining $q_0=0.538~{\rm GeV}$. 
Right panel: Dirac, obtaining $q_0=0.576~{\rm GeV}$. In the case of Dirac quark model, the global sign of hadronic matrix elements can not be known independently of the states phases: the convention is such that the discrepancy between Dirac and BT is minimal, $f_B>0$ and $f_{B^{*\prime}}>0$. }
\label{tab:ff_comp}
\end{table}

\section{Multihadronic states}

\begin{figure}[t]
\begin{center}
\begin{tabular}{cc}
	\includegraphics*[width=0.45\linewidth]{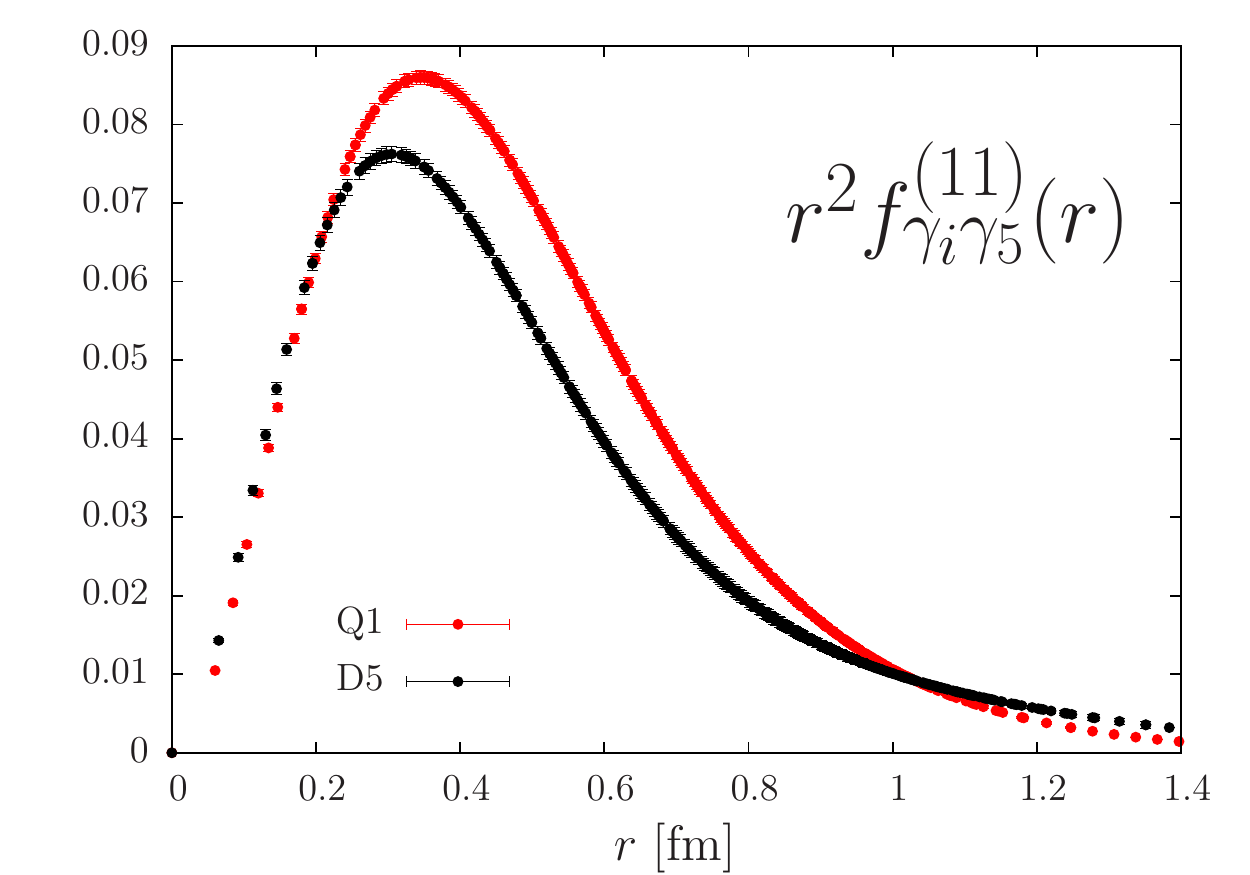}
&	\includegraphics*[width=0.45\linewidth]{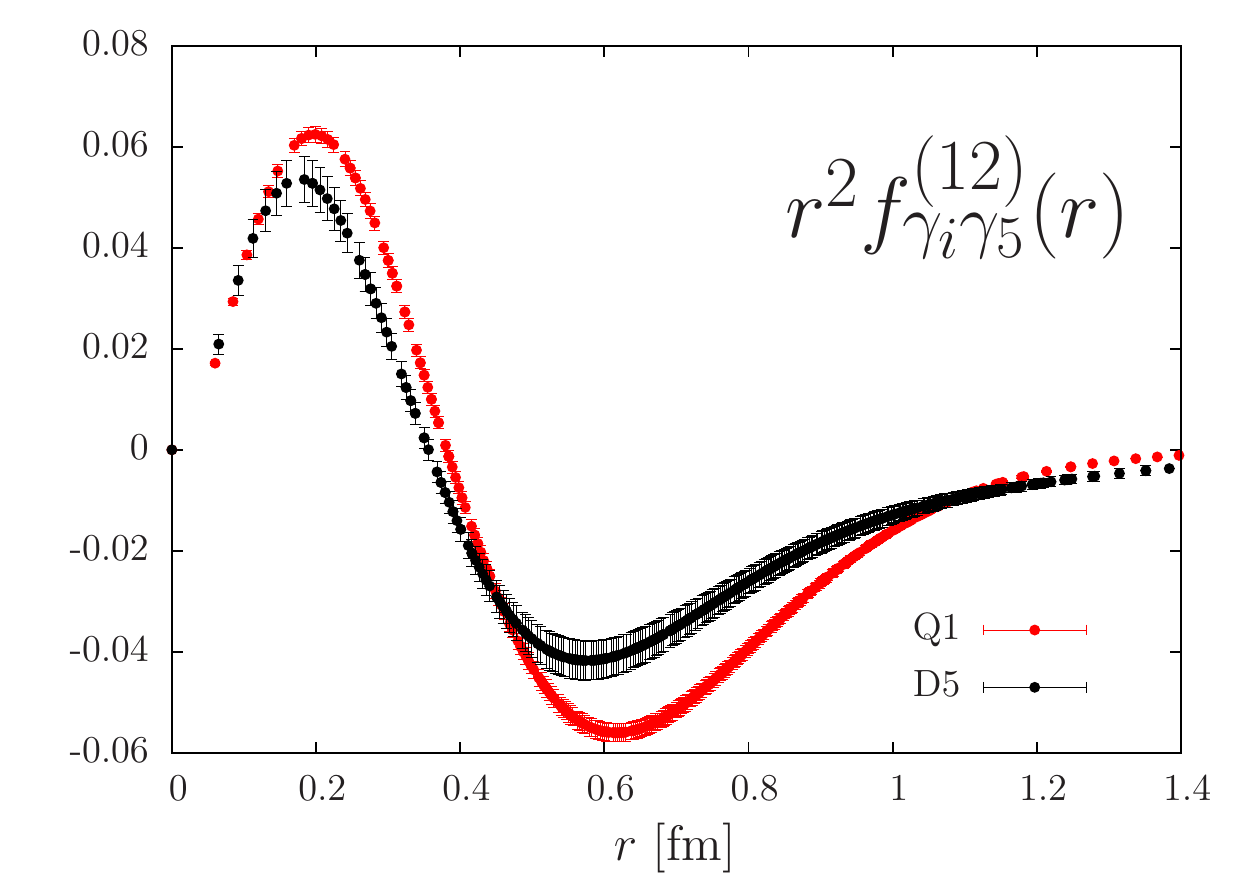}
\end{tabular}	
\end{center}
	\caption{Comparison of the renormalized distributions $r^2\, f^{(mn)}_{\gamma_i\gamma_5}(r)$ in physical units. The quenched result (ensemble Q1) is plotted in red and the dynamical case in black (ensemble D5).	
	\label{fig:quenched_cmp}}
\end{figure}

In many lattice studies, radial or orbital excitations of mesons lie near a multihadron threshold, making the extraction of excited states properties a difficult challenge. Interpolating operators that have a large overlap with a two-body system \cite{MichaelKW} are often used but they require more computer time and it is argued that bilinear interpolating operators are coupled only weakly with those states \cite{BarCE}. We have profited of our work to study that problem in details because it can be an unpleasant source of systematics. Within our lattice setup, the $B^{*\prime}$ radial excited vector meson lies near the multiparticle threshold $B_1^* \pi$ in $S$ wave where $B_1^*$ represents the axial $B$ meson, as can be seen in Table~\ref{tab:threshold}.  
\begin{table}[t]
	\begin{center}
\begin{tabular}{cc}
	\begin{tabular}{|c|c|c|c|c|}
\hline
	\ id\	&	$a\Sigma_{12}$	&	$a\delta$		&	$am_{\pi}$	&	$a\delta+am_{\pi}$		\\ 
\hline
	A5		&	$0.253(7)$		&	$0.155(4)$	&	$0.12625$	&	$0.281(4)$\\ 
	B6		&	$0.235(8)$		&	$0.141(4)$	&	$0.10732$	&	$0.248(4)$\\ 
	E5		&	$0.225(10)$		&	$0.133(6)$	&	$0.14543$	&	$0.278(6)$\\ 
	F6		&	$0.213(11)$		&	$0.129(3)$	&	$0.10362$	&	$0.233(3)$\\ 
	N6		&	$0.166(9)$		&	$0.092(3)$	&	$0.08371$	&	$0.176(3)$\\ 
\hline
	\end{tabular}
&
	\begin{tabular}{|c|c|c|}
\hline
	lattice	&	$m_\pi\,[{\rm MeV}]$	&	$r^{(12)}_n\,[{\rm fm}]$		\\ 
\hline
	A5		&	330		&	$0.369(13)$\\ 
	B6		&	280		&	$0.374(12)$\\ 
\hline
	E5		&	440		&	$0.369(11)$\\ 
	F6		&	310		&	$0.379(20)$\\ 
\hline
	N6		&	340		&	$0.365(12)$\\ 
\hline
	\end{tabular}
\end{tabular}
	\end{center}	
	\caption{Mass splittings $\Sigma_{12}=m_{B^{*\prime}}-m_B$ and $\delta = m_{B_1^{*}}-m_B$ for each lattice ensemble used to extrapolate $A^{(12)}_0(q^2)$ at the physical point (left); position of the node $r^{(12)}_n$ of the radial distribution $f^{(12)}_{\gamma_i\gamma_5}(r)$ on those ensembles (right).}
	\label{tab:threshold}
\end{table}
Assuming a non-interacting two-particle state, with the energy given by $E=m_{B_1^*}+m_{\pi}$, we are below (but near) threshold for all lattice ensembles used to get results at the physical point. Since our interpolating operators are coupled, in principle, to all states with the same quantum numbers, we could be sensitive to the $B^*_1 \pi$ state. However, if the coupling were not small, it would be difficult to interpret our $3 \times 3$ generalized eigenvalue problem (GEVP) results in our extraction of $g_{B^{*'}B\pi}$: we have seen a clear signal for the third excitation and it is far above the second energy level. Moreover, Fig.~\ref{fig:quenched_cmp} shows that the behaviour of density distributions $f^{(11)}_{\gamma_i \gamma_5}$ and $f^{(12)}_{\gamma_i \gamma_5}$ are similar at ${\rm N_f}=2$ and in the quenched approximation, while  the position of the node of the density distribution $f^{(12)}_{\gamma_i \gamma_5}$ is remarkably stable versus the pion mass, contrary to what would be expected in the case of a mixing with multiparticle states: results are collected in Table~\ref{tab:threshold}.
Finally, the qualitative agreement with quark models makes us confident that our measurement of the density distributions $f^{(12)}_\Gamma(r)$ probes transition amplitudes among $\bar{q}b$ bound states: in the quark model language they correspond to overlaps between wave functions.

\begin{figure}[t]
\begin{center}
\begin{tabular}{cc}
	\includegraphics*[width=0.4\linewidth]{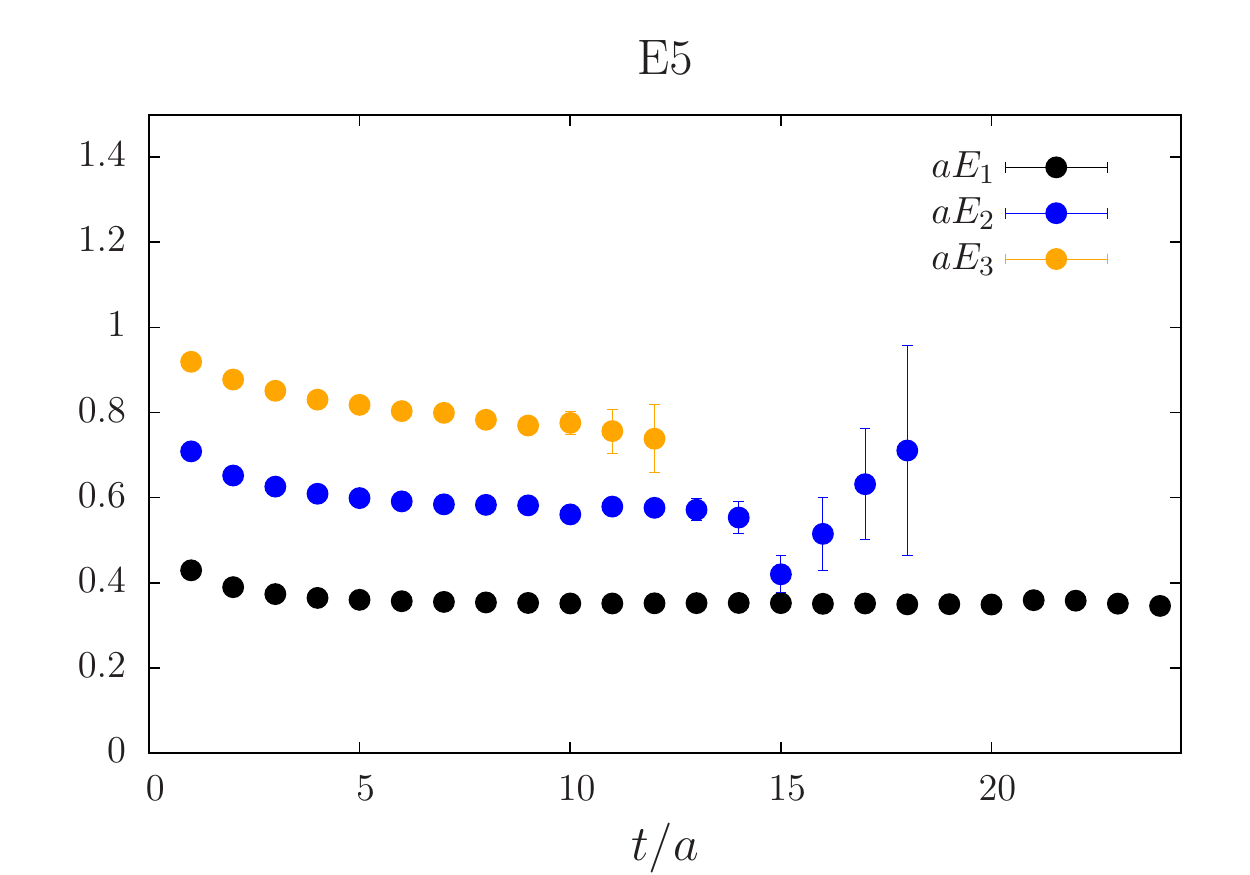}
&	\includegraphics*[width=0.4\linewidth]{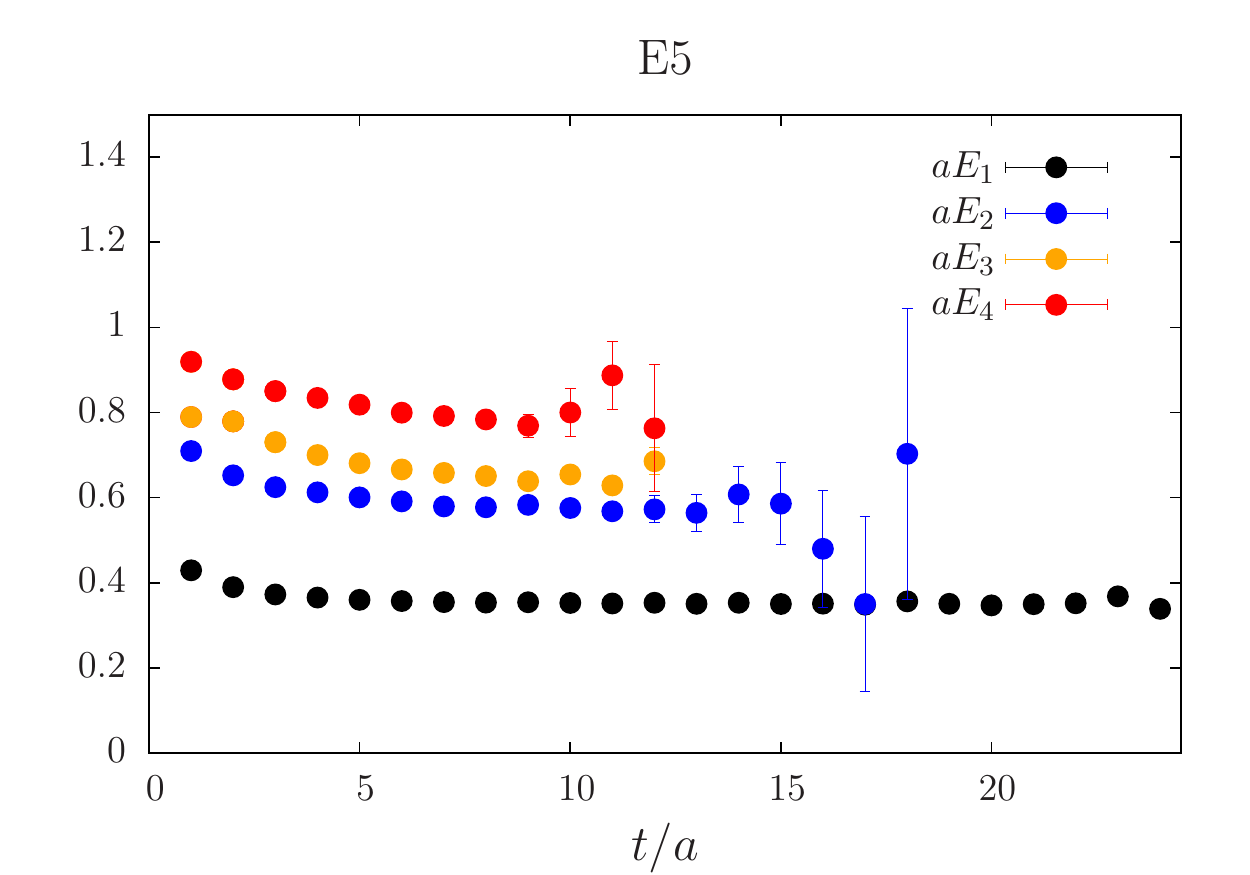}
\end{tabular}
\end{center}
	\caption{Effective mass plot extracted from a $3\times 3$ GEVP for the lattice ensemble E5 using $\bar{q} \gamma_k b$ interpolating operators (left); effective mass plot extracted from a $4\times 4$ GEVP for the lattice ensemble E5 using $\bar{q}\gamma_k b$ and $\bar{q} \nabla_k b$ interpolating operators (right).}	
\label{fig:GEVP_threshold}
\end{figure}

\begin{figure}[t]
\begin{center}
\begin{tabular}{cccc}
	\includegraphics*[width=0.2\linewidth,height=0.17\linewidth]{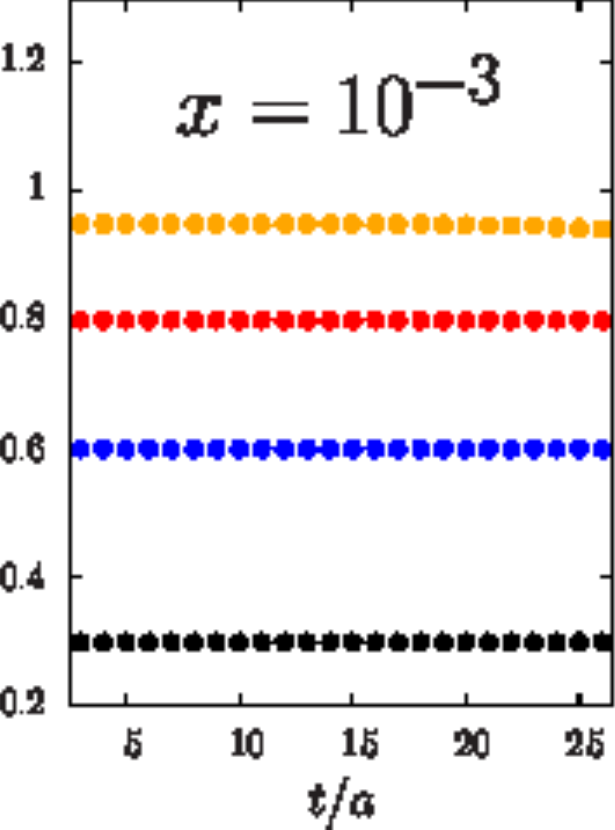}
& 	\includegraphics*[width=0.2\linewidth,height=0.17\linewidth]{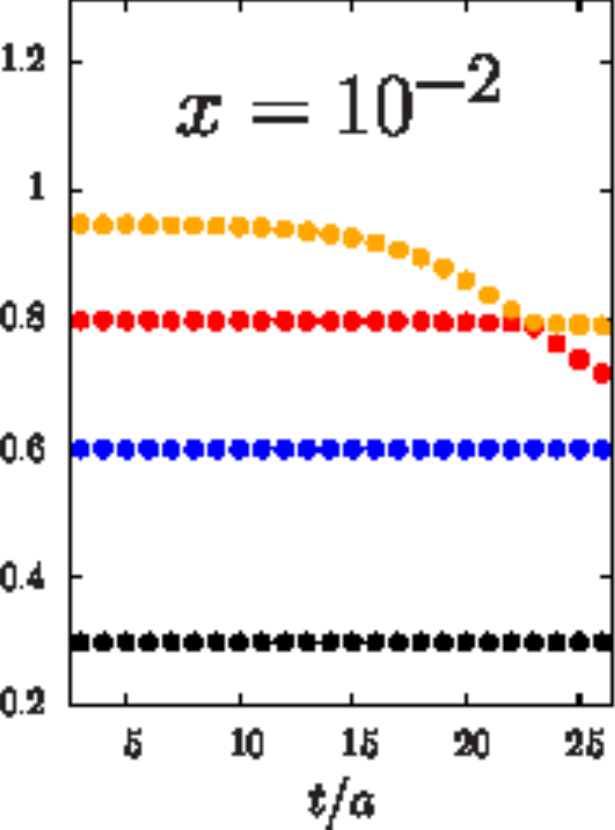}
&	\includegraphics*[width=0.2\linewidth,height=0.17\linewidth]{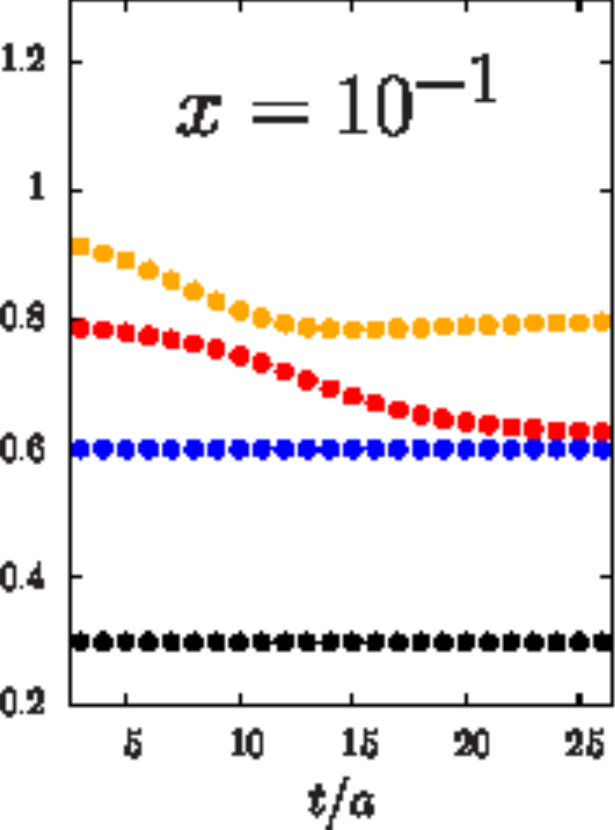}
&	\includegraphics*[width=0.2\linewidth,height=0.17\linewidth]{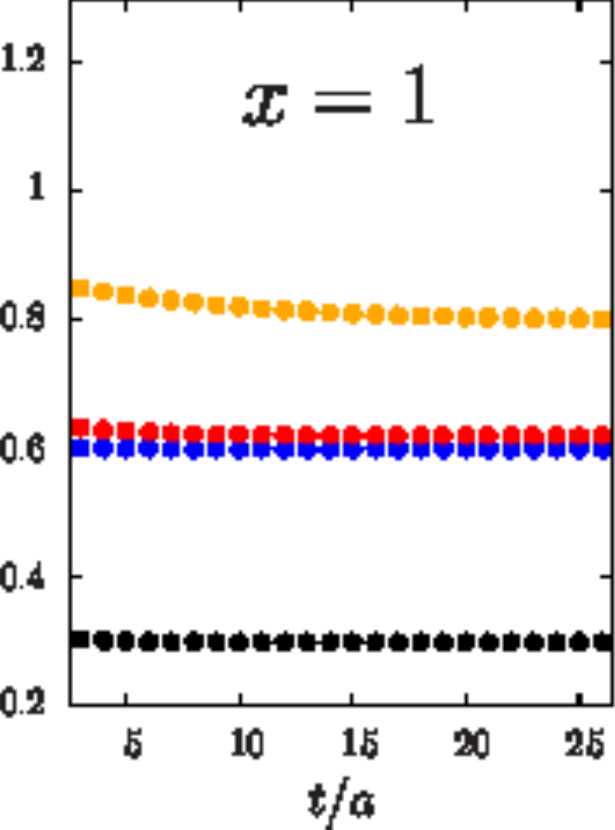}
\end{tabular}
\end{center}	
	\caption{\label{fig:plotgevpx} Effective energies of the two-point correlation function defined in our toy model,
	obtained by solving a $4\times 4$ GEVP for different values of $x$.}
\end{figure}

\begin{figure}[t]
\begin{center}
\begin{tabular}{cccc}
\includegraphics*[width=0.2\linewidth,height=0.16\linewidth]{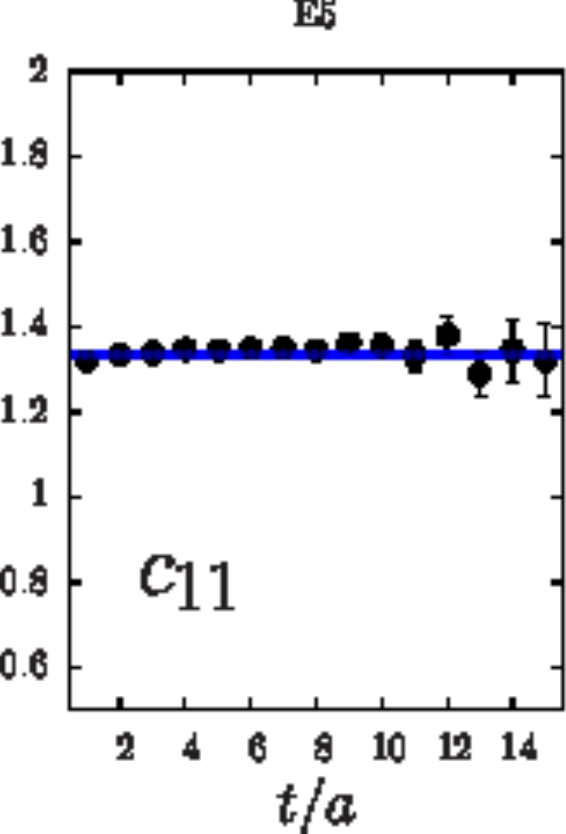}
&	\includegraphics*[width=0.2\linewidth,height=0.16\linewidth]{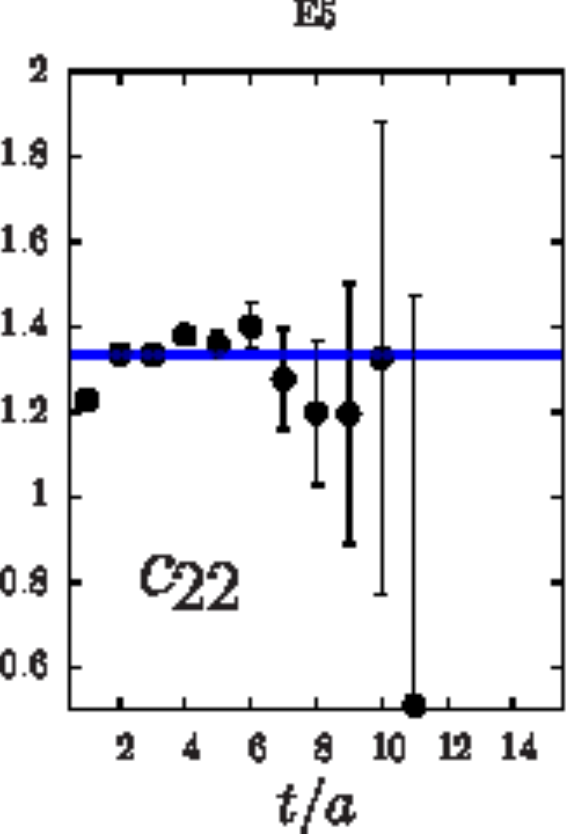}
&	\includegraphics*[width=0.2\linewidth,height=0.16\linewidth]{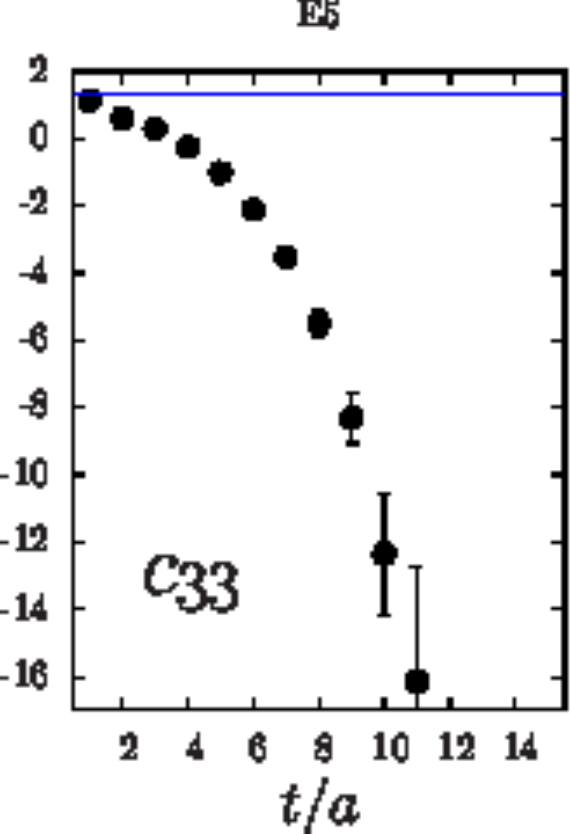}
&	\includegraphics*[width=0.2\linewidth,height=0.16\linewidth]{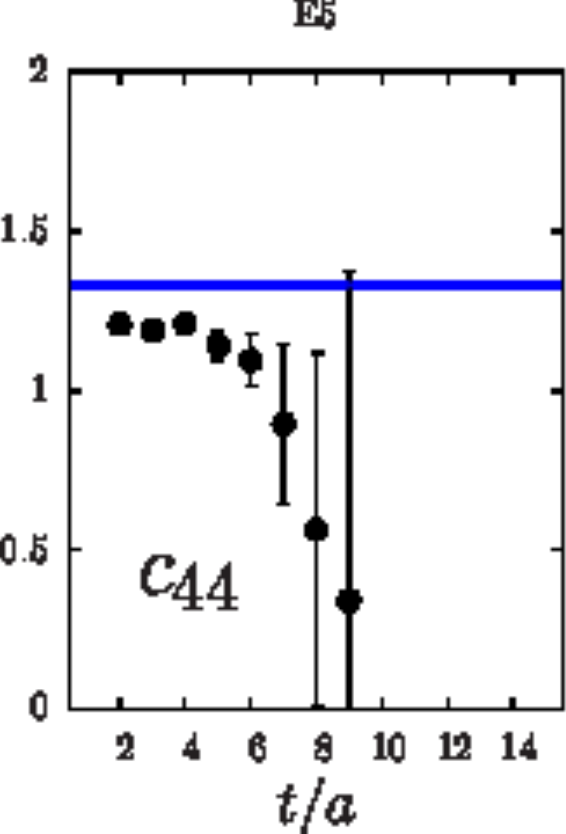}
\end{tabular}
\end{center}
\caption{\label{fig:chargesumnabla} Unrenormalized vector charge got from $f^{(nn)}_{\gamma_0}(r)$ on the lattice ensemble E5, using $\bar{q} \gamma_k h$ and $\bar{q} \nabla_k h$ interpolating operators. The blue line corresponds to the expected plateau using the nonperturbative estimate $Z_V = 0.750(5)$ extracted from~\cite{Fritzsch:2012wq}. }
\end{figure}

\begin{figure}[t]
\begin{center}
\begin{tabular}{c}
\begin{tabular}{ccc}
	\includegraphics*[height=0.2\linewidth]{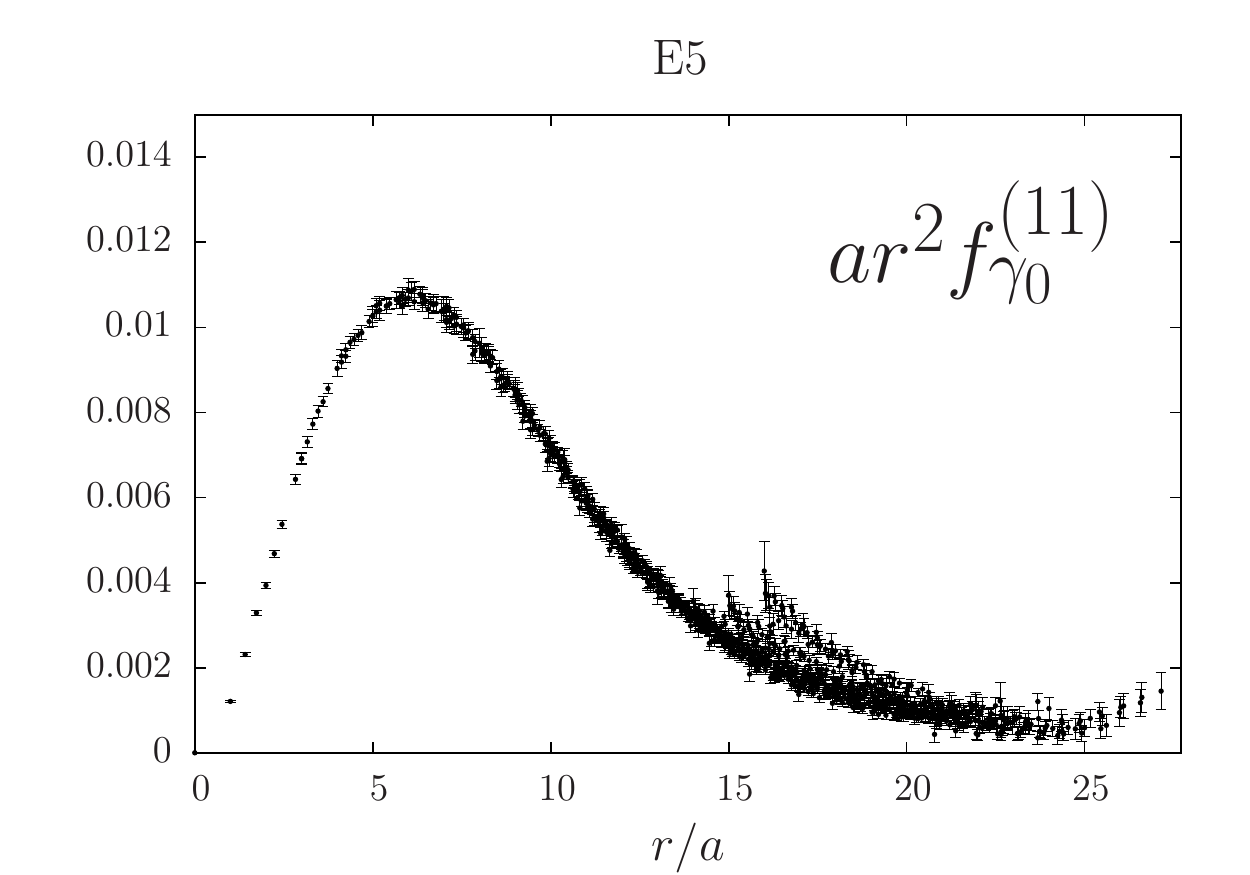}
&	\includegraphics*[height=0.2\linewidth]{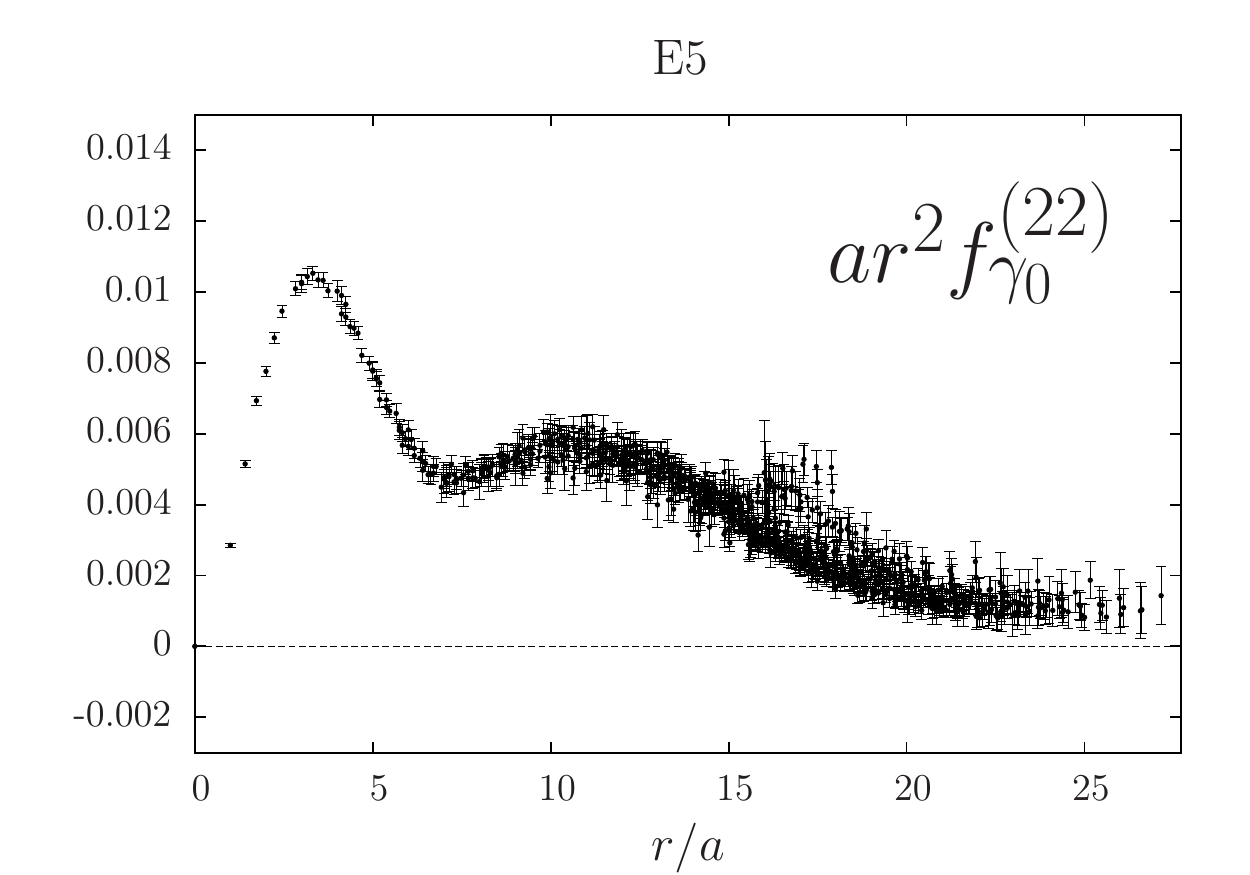}
&	\includegraphics*[height=0.2\linewidth]{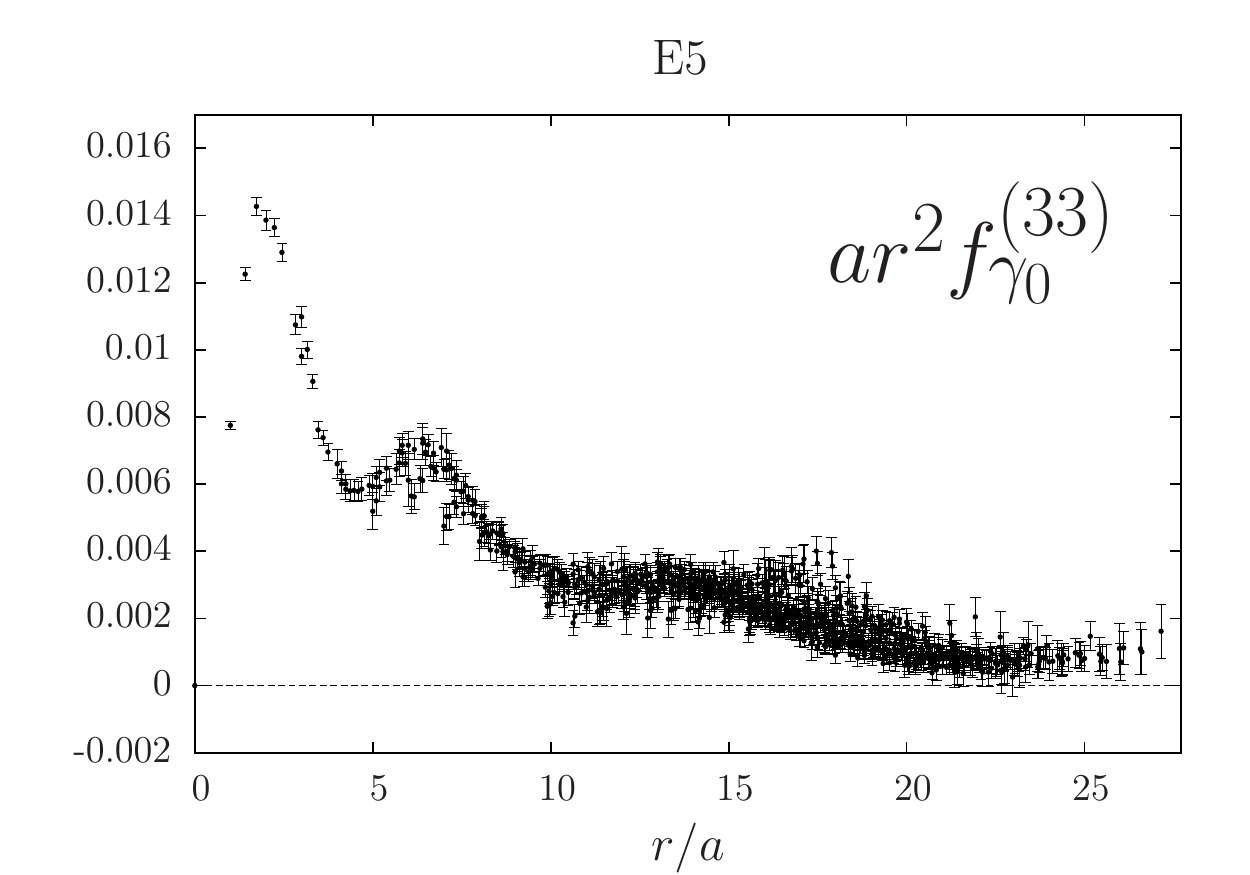}
\end{tabular}
\\
\begin{tabular}{ccc}
	\includegraphics*[height=0.2\linewidth]{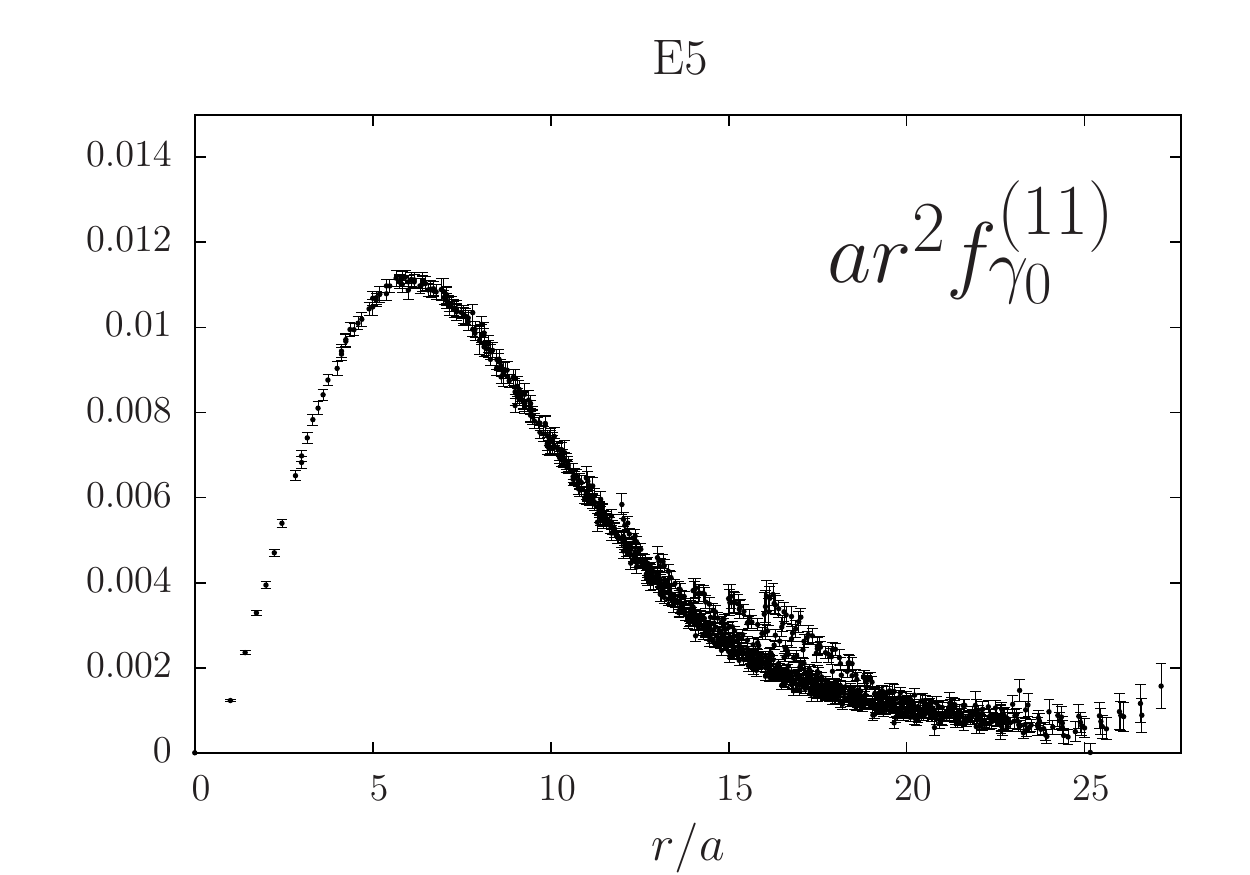}
&	\includegraphics*[height=0.2\linewidth]{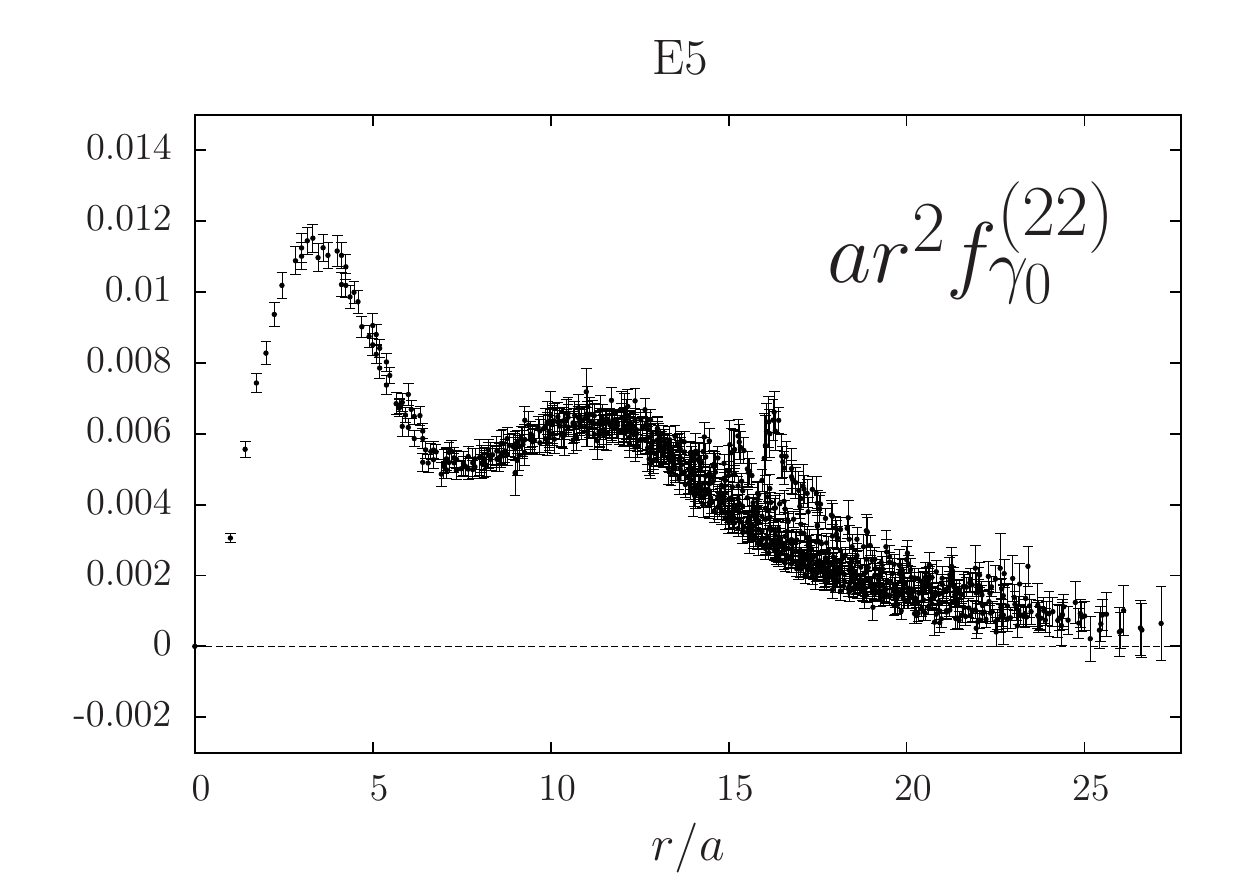} 
&	\includegraphics*[height=0.2\linewidth]{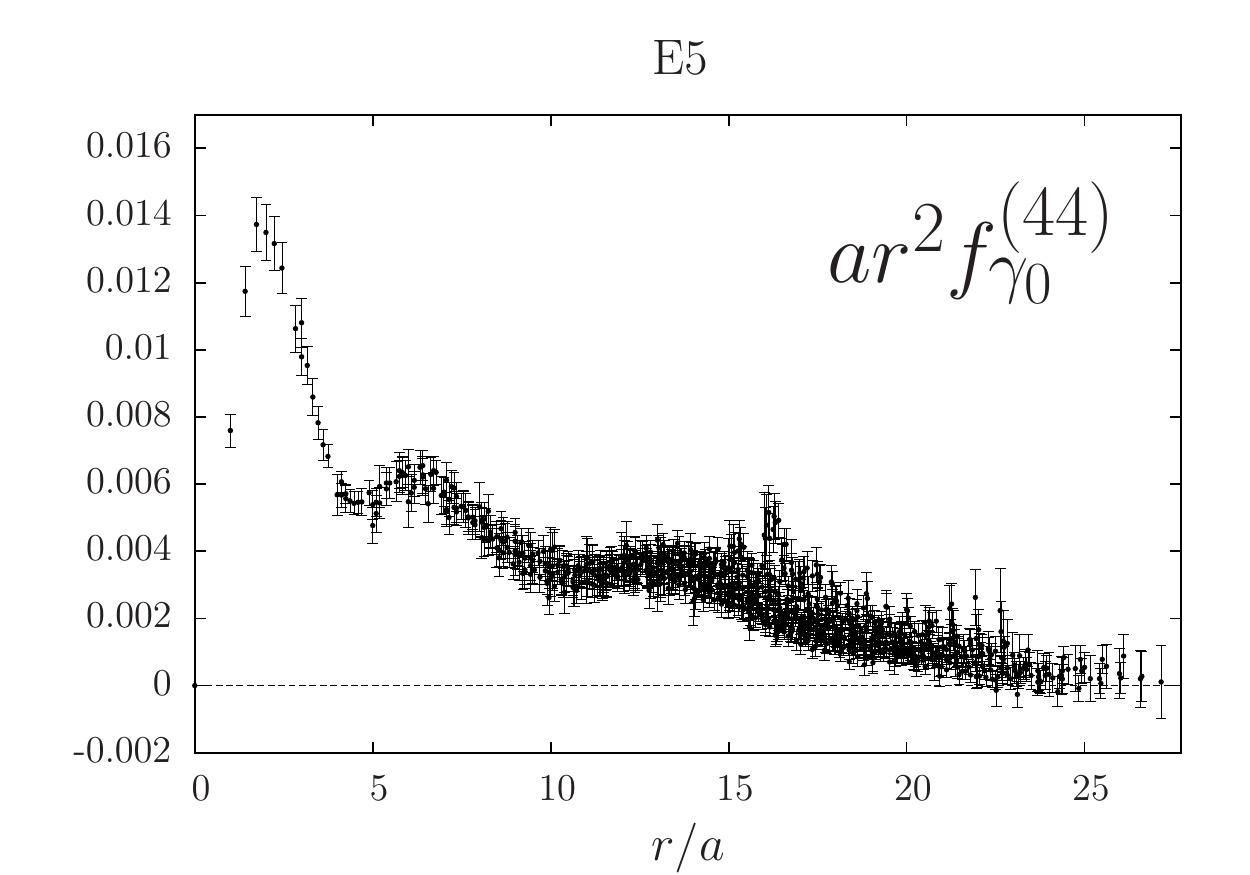}
\end{tabular}
\end{tabular}
\end{center}
\caption{\label{fig:densitycharge} Density distributions $ar^2f^{(nn)}_{\gamma_0}(r/a)$, $n=1,2,3$ (top) and $n=1,2,4$ (bottom) on the lattice ensemble E5, using only $\bar{q} \gamma_k h$ (top) and including $\bar{q} \nabla_k h$ interpolating fields (bottom) in the analysis.}
\end{figure}

The picture does change if, in addition to the Gaussian smearing operators $\mathcal{V}^{(i)}_{k}(x) =  \overline{u}^{(i)}(x) \gamma_k h(x)$ used so far, we insert a second kind of interpolating operators which could couple to the two-particle state: $\mathcal{V}^{(i)}_{k}(x) =  \overline{u}^{(i)}(x) \overleftarrow{\nabla}_k h(x)$. As can be seen in Fig.~\ref{fig:GEVP_threshold}, the GEVP indeed isolates a new state, slightly above the radial excitation of the vector meson, whose interpretation can be guessed from Table~\ref{tab:threshold}. The effective mass of the ground state and first excited state remain unchanged, as we indicate in Table~\ref{tab:eff_mass_multihadron}.
\begin{table}[t]
\begin{center}
\begin{tabular}{|l|l|c|c|c|}
\cline{3-5}
\multicolumn{1}{c}{}&\multicolumn{1}{c}{}	&\multicolumn{1}{|c}{}	$a\Sigma_{12}$		&	$a\Sigma_{13}$		&	$a\Sigma_{14}$  \\
\hline 
	\multirow{2}{*}{E5}	&	$\gamma_k$			&	$0.225(8)$		&	$0.417(21)$	&  	$\times$	\\  
\cline{2-5}
		&	$\gamma_k, \nabla_k$	&	$0.218(12)$	&	$0.278(17)$	&  	$0.422(12)$	 	\\  
\hline
	\multirow{2}{*}{A5}	&	$\gamma_k$			&	$0.257(6)$		&	$0.467(23)$	&  	$\times$	\\  
\cline{2-5}
		&	$\gamma_k, \nabla_k$	&	$0.254(7)$	&	$0.315(11)$	&  	$0.459(24)$	 	\\  
\hline
 \end{tabular} 
\end{center}
\caption{Energy levels extracted from the GEVP (ensembles E5 and A5). In the first raw only Gaussian smeared operators $\mathcal{V}^{(i)}_{k}(x) =  \overline{u}^{(i)}(x) \gamma_k h(x)$ are used. In the second raw, both interpolating operators of the form $\mathcal{V}^{(i)}_{k}(x) =  \overline{u}^{(i)}(x) \gamma_k h(x)$ and $\mathcal{V}^{(i)}_{k}(x) =  \overline{u}^{(i)}(x) \protect\overleftarrow{\nabla}_k h(x)$.} 
\label{tab:eff_mass_multihadron}
\end{table}
To try to understand this fact, we have performed a test on a toy model. The spectrum contains five states, with energies $E^{(i)}=\{0.3, 0.6, 0.63, 0.8, 0.95\}$. The $1^{\rm st}$ and $2^{\rm nd}$ excited states are almost degenerate. With a basis of five interpolating fields, the matrix of couplings reads: 
\begin{equation}
M^{x}=\left[ \begin{array}{ccccc}
0.60&0.25&x \times 0.40&0.10&0.50\\
0.61&0.27&x \times 0.39&0.11&0.51\\
0.58&0.24&x \times 0.42&0.12&0.52\\
0.57&0.25&x \times 0.41&0.10&0.49\\
0.56&0.26&x \times 0.36&0.08&0.48\\
\end{array}
\right],
\end{equation}
where $x$ can be varied from $10^{-3}$ (third interpolating field almost not coupled to the spectrum under investigation) to $1$ (third interpolating field as strongly coupled to the spectrum as the other operators). We solve a GEVP on the $4 \times 4$ matrix of correlators $C^x_{ij}$ defined by
$C^x_{ij}(t)=\sum_{n=1}^5 M^x_{ni}M^x_{nj} e^{-E_n t}$.
We show in Fig.~\ref{fig:plotgevpx} effective masses obtained from the generalized eigenvalues, when $x$ is growing. A transition is clear: the GEVP isolates the states 1, 2, 4 and 5 at very small $x$ and then, as $x$ is made larger, the states 1, 2, 3 and 4. In other words, GEVP can ``miss" an intermediate state of the spectrum if, by accident, the coupling of the interpolating fields to that state is suppressed. Our claim is that, using interpolating fields $\bar{q} \gamma_i h$, we have no chance to couple to multi-hadron states while inserting an operator $\bar{q} \nabla_i h$ could isolate the $B^*_1 \pi$ two-particle state.\\
\noindent We have also examined the radial distribution of the vector density, because the conservation of the vector charge is an excellent indicator of a possible source of uncontrolled systematics if it is strongly violated. 
It is defined similarly to the axial density distribution by replacing the axial density with $\mathcal{O}_{\Gamma} = \overline{\psi}_l  \gamma_0 \psi_l$. With the interpolating field $\bar{q}\nabla_k h$ included in the basis, together with $\bar{q}\gamma_k h$, we show in Fig. \ref{fig:chargesumnabla} the ``effective" charge density distributions $f^{(nn)}_{\gamma_0}(r)$ integrated over $r$, in function of the time $t$ entering the (summed) GEVP. In the cases of $f^{(11)}_{\gamma_0}(r)$ and $f^{(22)}_{\gamma_0}(r)$, plateaus are clearly compatible with $1/Z_V$, where $Z_V$ is the renormalization constant of the vector current extracted from \cite{Fritzsch:2012wq}, while, for $f^{(33)}_{\gamma_0}(r)$, we observe a divergence with time. Concerning $f^{(44)}_{\gamma_0}(r)$, a (very short) plateau shows up again around $1/Z_V$. Once more, the main lesson is that the second excited state isolated by the GEVP is hard to interpret as a $\bar{q}b$ bound state whereas the first excited state is. Density distributions themselves are showed in Fig.~\ref{fig:densitycharge}. Plots on the top correspond to the basis with only $\bar{q}\gamma_i h$-kind interpolating fields of the $B^*$ meson and those on the bottom are obtained after incorporating $\bar{q}\nabla_k h$-kind in the analysis. We note similar facts as for the spectrum: $f^{(11)}_{\gamma_0}(r)$ and $f^{22}_{\gamma_0}(r)$ are almost the same, $f^{(33)}_{\gamma_0}(r)$ of the top looks like $f^{(44)}_{\gamma_0}(r)$ on the bottom. Finally it revealed impossible to obtain a stable density for $f^{(33)}_{\gamma_0}(r)$ when we include $\bar{q} \nabla_k h$ operators in the analysis. Actually, it is just a rephrasing of the observation made just above. 


\begin{thebibliography}{99}

\bibitem{BecirevicVP}
  D.~Becirevic, J.~Charles, A.~LeYaouanc, L.~Oliver, O.~P\`ene and J.~C.~Raynal,
JHEP {\bf 0301}, 009 (2003).

\bibitem{OhkiPY}
H.~Ohki, H.~Matsufuru, and T.~Onogi,
Phys. Rev. {\bf D77}, 094509 (2008); D.~Becirevic, B.~Blossier, E.~Chang, and B.~Haas,
Phys. Lett. {\bf B679}, 231 (2009); W.~Detmold, C.~D. Lin, and S.~Meinel, Phys. Rev. {\bf D85}, 114508 (2012);
D.~Becirevic and F.~Sanfilippo, Phys. Lett. {\bf B721}, 94 (2013); F.~Bernardoni {\it et al.} [ALPHA Collaboration],
Phys. Lett. {\bf B740}, 278 (2015).

\bibitem{GodangIM}
  R.~Godang,
PoS ICHEP {\bf 2012}, 330 (2013).

\bibitem{KhodjamirianHB}
A.~Khodjamirian, R.~Ruckl, S.~Weinzierl, and O.~I. Yakovlev,
Phys. Lett. {\bf B457}, 245 (1999).

\bibitem{BlossierQMA}
  B.~Blossier, J.~Bulava, M.~Donnellan and A.~G\'erardin,
Phys. Rev.  {\bf D87}, no. 9, 094518 (2013); B.~Blossier and A. G\'erardin, Phys. Rev. {\bf D94}, 074504 (2016).

\bibitem{LeYaouancprivate} A.~Le~Yaouanc, private communication.

\bibitem{MichaelKW}
  C.~Michael,
PoS LAT {\bf 2005}, 008 (2006).

\bibitem{BarCE}
  O.~B\"ar and M.~Golterman,
Phys. Rev. {\bf D87}, no. 1, 014505 (2013).

\bibitem{Fritzsch:2012wq}
  P.~Fritzsch, F.~Knechtli, B.~Leder, M.~Marinkovic, S.~Schaefer, R.~Sommer and F.~Virotta,
  Nucl. Phys. {\bf B865}, 397 (2012).

\end{thebibliography}
\end{document}